\documentclass[aps,prb,twocolumn,showpacs,floatfix,longbibliography]{revtex4-1} 
\usepackage{amsfonts}
\usepackage{graphicx}
\usepackage{amsmath}
\usepackage{times}
\usepackage{amssymb}
\usepackage{color}
\usepackage{subfigure}
\usepackage[colorlinks,bookmarks=false,citecolor=blue, linkcolor=red, urlcolor=magenta]{hyperref}

\newcommand{\diagram}[1]{\vcenter{\hbox{\includegraphics[scale=0.45]{./#1.pdf}}}}

\begin{document}

\title{Numerical continuum tensor networks in two dimensions}
\author{Reza Haghshenas}
\email{haqshena@caltech.edu}
\author{Zhi-Hao Cui}
\author{Garnet Kin-Lic Chan}
\email{garnetc@caltech.edu}

\affiliation{Division of Chemistry and Chemical Engineering, California Institute of Technology, Pasadena, California 91125, USA}

\begin{abstract}
  We describe the use of tensor networks to numerically determine wave functions of interacting two-dimensional fermionic models in the continuum limit. We use two different tensor network states: one based on the numerical continuum limit of fermionic projected entangled pair states obtained via a tensor network formulation of multi-grid, and another based on the combination of the fermionic projected entangled pair state with layers of isometric coarse-graining transformations. We first benchmark our approach on the two-dimensional free Fermi gas then proceed to study the two-dimensional   interacting  Fermi gas with an attractive interaction in the unitary limit, using tensor networks on grids with up to 1000 sites. 
\end{abstract}
\maketitle

\section{Introduction}

Understanding the collective behavior of quantum many-body systems is a central theme in physics.
While it is often discussed using lattice models, there are
systems where a continuum description is essential. 
One such case is found in superfluids,\cite{ANDERSON1973153} 
where recent progress in precise experiments on ultracold atomic Fermi gases has opened up new opportunities 
to probe key aspects of the phases.\cite{Bloch:2008}  On the theoretical side, this requires solving a continuum fermionic quantum many-body problem. 
For example, various quantum Monte Carlo (QMC) methods have been applied to study
the cross-over from Bardeen-Cooper-Schrieffer superfluidity to Bose-Einstein condensation in two-dimensional Fermi gases.\cite{Bertaina:2011, Shi:2015, Galea:2016, Rammelm:2016} However, the applicability of (unbiased) quantum Monte Carlo is restricted to 
special parameter regimes due to the fermion sign problem.\cite{Troyer:2005} Thus, devising numerical methods that 
can address general continuum quantum many-body physics remains an important objective.

Tensor network states (TNS) are classes of variational states that have become widely used in quantum lattice models.
They are complementary to QMC methods as TNS algorithms are typically formulated
without incurring a sign problem.
In 1D, matrix product states (MPS) now provide almost exact numerical results via the DMRG algorithm.\cite{White:1992}
In 2D, reaching a similar level of success has been harder, but
much progress has been made using projected entangled pair states (PEPS),\cite{Murg:2009} which generalize MPS to higher dimensions
in a natural fashion.
PEPS calculations now provide accurate results for a broad range of quantum lattice problems\cite{Corboz:2011, Corboz:2013, zheng2017stripe, Haghshenas:2018May, Haghshenas:2019May} and there have been many developments to extend the range of the techniques,
for example to  long-range Hamiltonians,\cite{Matthew:2018, Li2019, Matthew2020} thermal states,\cite{Czarnik:2012}  and real-time dynamics.\cite{Czarnik:2018} Also, much work has been devoted to improving
the numerical efficiency and stability of PEPS computations.\cite{Corboz:2016, Vanderstraeten:2016, Haghshenas:2019, Dong:2019, Liao:2019}

Formulating tensor network states and the associated algorithms in the continuum remains a challenge.
In 1D, so-called continuous MPS~\cite{Verstraete:2010} provide an analytical ansatz in the continuum 
and have been applied to several problems, including 1D interacting bosons/fermions and quantum field theories.~\cite{Haegeman:2010, Ganahl2018, Chung:2015, Ganahl:2017} Alternatively, the continuum description can be reached by taking the numerical limit of a set of tensor network states formulated on lattices with a discretization parameter $\epsilon$, for $\epsilon\to 0$. This kind of numerical continuum MPS calculation has also been demonstrated in conjunction with a variety of optimization algorithms.~\cite{Stoudenmire2012, Dolfi:2012, Stoudenmire2017}

In two dimensions, despite several proposals,~\cite{Verstraete:2010, Haegeman:2013, Dolfi:2012, Tilloy:2019} the appropriate analytical form of the continuum PEPS ansatz remains unclear. In this work, we carry out continuum tensor network calculations in 2D by taking the numerical limit of a lattice discretization parameter. We explore two types of ansatz to approach the continuum. The first uses the
numerical continuum limit of the lattice fermionic PEPS. Here, to connect the lattice PEPS at different scales when taking this limit and to ensure an
efficient optimization on finer scales we use a multi-grid like algorithm (a generalization of the MPS multigrid algorithm). The second is based on a combination of fermionic PEPS with a tree of isometries
that successively coarse grains the continuum into discrete lattices. 
Using these 2D numerical continuum tensor network states, we demonstrate
how we can study fermionic physics in the continuum limit, applying the ansatz both
to the challenging (for tensor networks) case of the free Fermi gas, as well as
the attractive interacting Fermi gas in the unitary limit that can be realized in ultracold atom experiments. 
In the first case, we can benchmark against exact results, while in the second
we can perform direct comparisons of the tensor network results to recent
QMC calculations at half-filling (where there is no sign problem) in the continuum and thermodynamic limits.

The remainder of the paper is organized as follows. We first introduce the fermionic continuum Hamiltonian
of interest in Sec.~\ref{Sec:continuumHamiltonian} and describe how to discretize it in a manner consistent with open boundary conditions in Sec.~\ref{Sec:DiscritizationMethod}. The two types of fermionic tensor network states are discussed
in Sec.~\ref{Sec:fermionic-tensor-network-ansatz} along with the optimization algorithms used for them. 
 We then present our numerical benchmarks for the free and interacting Fermi gases in Sec.~\ref{Sec:BenchmarkResults}. We summarize our work in Sec.~\ref{Sec:CONCLUSION} and discuss future research directions.

\begin{figure}
\begin{center}
\includegraphics[width=1.0 \linewidth]{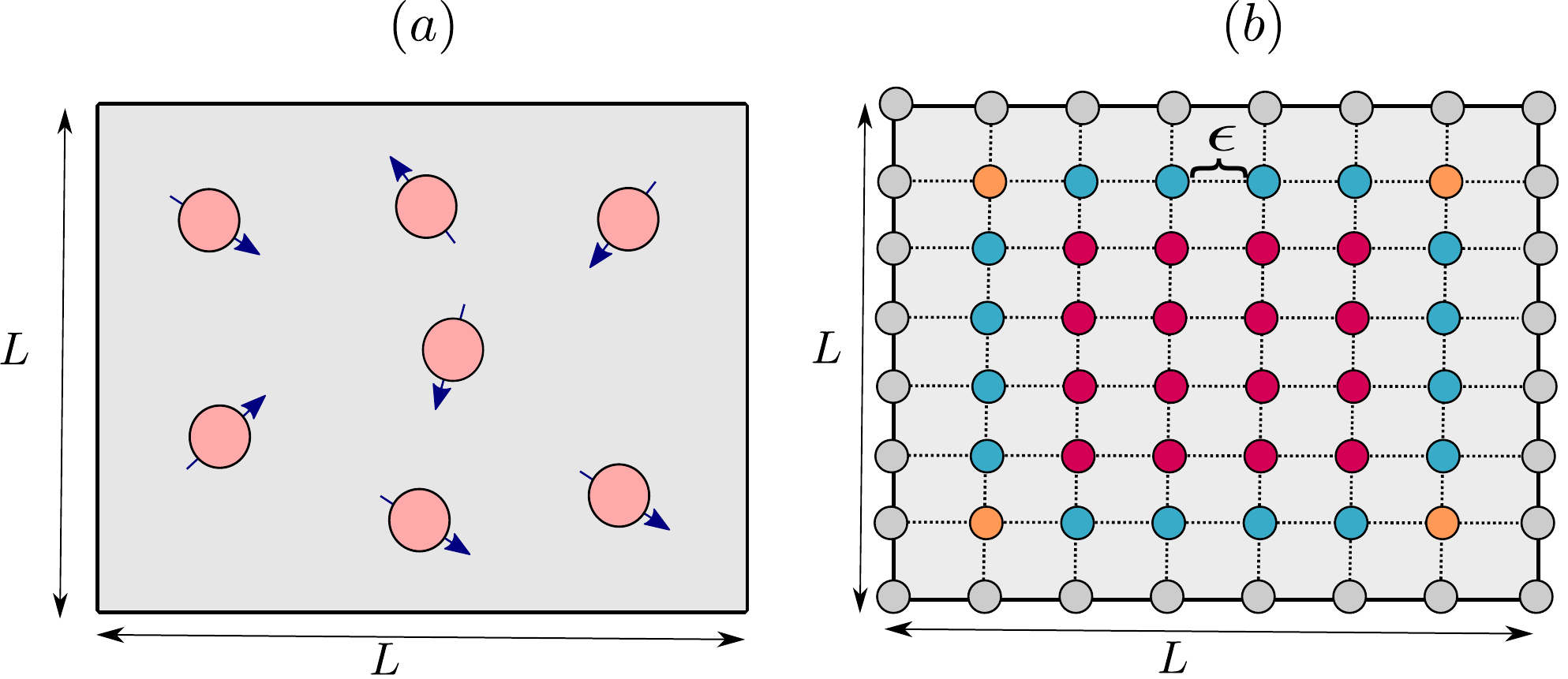} 
  \caption{(Color online) (a) A two-dimensional square $L \times L$ box containing spinful fermions. The fermions are confined to the box
by an infinite potential at the walls of the box. (b) Lattice discretization of the $L \times L$ box into $8 \times 8$ grid points. 
Using Dirichlet boundary conditions, the wave function is zero at the gray points (boundary points). The active points
where the wave function takes non-trivial values are defined on the $6 \times 6$ lattice, with a lattice spacing of $\epsilon=\frac{L}{7}$. 
The grid points with blue and orange colors require special treatment in the fourth-order finite difference approximation, 
in order to provide an accurate discretization. }
  \label{fig:Lattice}
\end{center}
\end{figure}

\section{Interacting Fermi gas}
\label{Sec:continuumHamiltonian}

For concreteness, it is useful to define a particular continuum model. In this work
 we will consider the free and interacting Fermi gases in two dimensions. The Hamiltonian is given by
\begin{eqnarray}\label{eq:Ham}
  \mathcal{H} =&&\sum_{\sigma=\uparrow, \downarrow}  \int d\textbf{r} \,  \psi_{\sigma}^\dagger(\textbf{r}) (-\frac{1}{2} \nabla^{2} -\mu)\psi_{\sigma}(\textbf{r})
 \nonumber \\
&&  +g \int \! \! \int d\textbf{r} d\textbf{r}' \psi_{\uparrow}^\dagger(\textbf{r}) \psi_{\uparrow}(\textbf{r})  \delta(\textbf{r}-\textbf{r}') \psi_{\downarrow}^\dagger(\textbf{r}') \psi_{\downarrow}(\textbf{r}')
\end{eqnarray}
where $\psi_{\sigma}^\dag(\textbf{r})$ and $\psi_{\sigma}(\textbf{r})$ are fermionic field operators, creating and annihilating a fermion with spin $\sigma$ at position $\textbf{r}$, respectively. The fermionic field operators satisfy the anticommutation  relation $\{ \psi_{\sigma}^\dagger(\textbf{r}),  \psi_{\sigma'}(\textbf{r}') \}=\delta(\textbf{r}-\textbf{r}')\delta_{\sigma \sigma'}$. The coupling parameters $\mu$ and $g$ denote the chemical potential
(which controls the number of particles) and strength of interaction in the system. We assume the system is confined in a $L \times L$ square box ($0<r_x, r_y<L$), so that the potential outside the box is $V(\textbf{r})=\infty$. When $g=0$, we have a free fermion gas confined to the box.

\section{Hamiltonian lattice discretization}
\label{Sec:DiscritizationMethod}

Because we define the continuum properties as a numerical limit, we need to
first discretize the continuum Hamiltonian $\mathcal{H}$. To do so, we replace 
the continuum space $L \times L$ box by a lattice containing $(N+1) \times (N+1)$ grid points with lattice spacing $\epsilon=\frac{L}{N}$.
Due to the Dirichlet (open) boundary conditions, the wave function is zero on the boundary of the grid. Thus the non-trivial part of the quantum state is defined on $(N-1) \times (N-1)$ grid points (we refer to these as the ``active points''). The lattice discretization is shown in Figs.~\ref{fig:Lattice}(a, b).

The kinetic energy operator can be represented using a finite difference stencil on the grid. Using $2$nd and $4$th order finite difference approximations, the Laplacian operator $\int d\textbf{r} \,  \psi_{\sigma}^\dagger(\textbf{r}) \nabla^{2} \psi_{\sigma}(\textbf{r})$
is replaced respectively by
\begin{eqnarray}
 && 
\text{2nd order} \ \rightarrow \frac{1}{\epsilon^{2}} \sum_{\langle ij\rangle,\sigma} c^{\dagger}_{i\sigma}c_{j\sigma}-\frac{4}{\epsilon^{2}}\sum_{i\sigma}c^{\dagger}_{i\sigma}c_{i\sigma}+\mathcal{O}(\epsilon^{2}) \nonumber \\
&&  \text{4th order} \rightarrow \frac{16}{12\epsilon^{2}}\sum_{\langle ij\rangle,\sigma} c^{\dagger}_{i\sigma}c_{j\sigma}-\frac{1}{12\epsilon^{2}}\sum_{\langle \langle \langle ij\rangle \rangle \rangle,\sigma} c^{\dagger}_{i\sigma}c_{j\sigma} \nonumber \\&&
-\frac{60}{12\epsilon^{2}}\sum_{i\sigma}c^{\dagger}_{i\sigma}c_{i\sigma}+\mathcal{O}(\epsilon^{4}) \label{eq:discrete}
\end{eqnarray}
where $c, c^{\dagger}$ are fermionic lattice operators and the symbols $\langle ij \rangle$ and  $\langle \langle \langle ij\rangle \rangle \rangle$
denote nearest neighbor and  third nearest neighbor pairs.


Because the wave function is not smooth at the edge of the box, 
some care must be taken in applying the stencils to  ensure that
boundary errors of lower order in $\epsilon$ than implied by the stencil formula do not appear.
In the $2$nd order approximation, the second-derivative takes the form
$\sim \frac{-4\psi_0 + \psi_1 + \psi_{-1}}{\epsilon^2}$ where the indices on $\psi$ denote the $x$ coordinate. Taking the index $-1$ to refer
to the left boundary, we see that representing the wave function on only the interior $N-1$ active points, while using the
spacing $\epsilon = 1/(N+1)$ (rather than the naive spacing of $\epsilon=1/N$) is consistent with choosing the boundary condition $\psi_{-1}=0$ (and similarly for the right boundary). Using this definition of $\epsilon=1/(N+1)$ we obtain
the full $2$nd order lattice discretized Hamiltonian as
\begin{eqnarray*}
 \mathcal{H}^{(2)}_{\epsilon}&&=\frac{1}{\epsilon^{2}}\sum_{\langle ij\rangle,\sigma} c^{\dagger}_{i\sigma}c_{j\sigma}+h.c.-\frac{4}{\epsilon^{2}}\sum_{i\sigma}c^{\dagger}_{i\sigma}c_{i\sigma}\\
 &&+\bar{g}\sum_{i} c^{\dagger}_{i\uparrow}c_{i\uparrow}c^{\dagger}_{i\downarrow}c_{i\downarrow}-\mu \sum_{i\sigma}  c^{\dagger}_{i\sigma}c_{i\sigma}+\mathcal{O}(\epsilon^{2}),
\end{eqnarray*}
where $\bar{g}$ is a regularized $\delta$ function interaction parameter, whose regularization procedure
is described in detail in Secs.~\ref{sec:interactinggas} and~\ref{sec:appendixb}.

\begin{figure}
\begin{center}
\includegraphics[width=1.0 \linewidth]{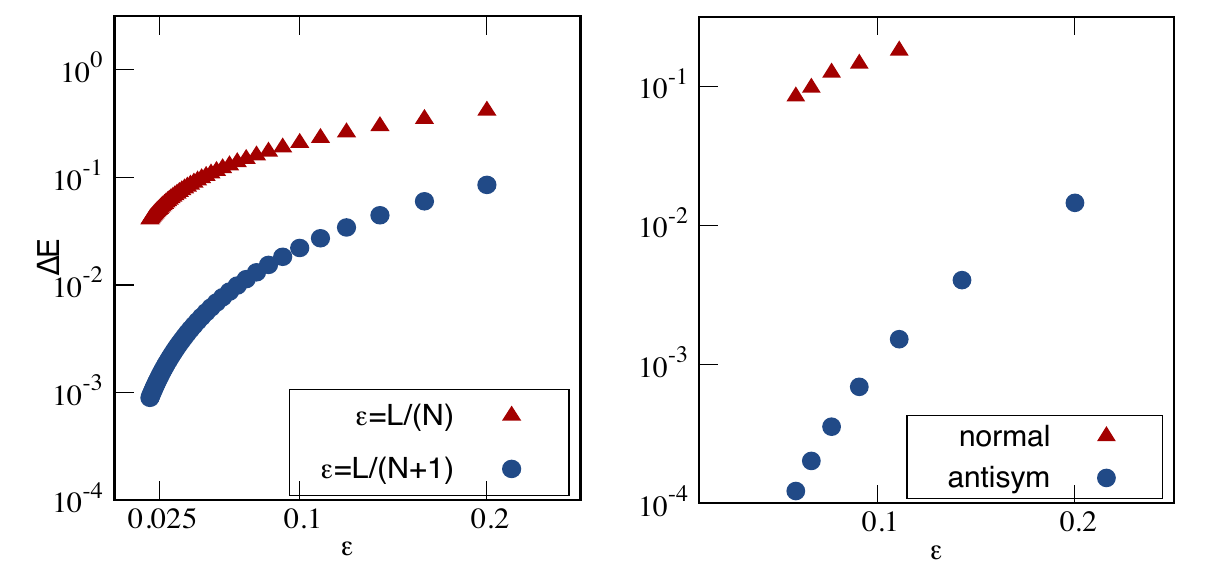} 
  \caption{(Color online) Relative error of ground-state energy of two fermions in a $1 \times 1$ box as a function of lattice spacing $\epsilon$. (a) Using the lattice spacing of $\epsilon=1/(N+1)$ (consistent with the Dirichlet boundary results) leads to the correct $2$nd order convergence with $\epsilon$ using the $2$nd order discretization formula. (b) Assuming an antisymmetric continuation of the wave function past the boundary (antisym) in the $4$th order discretization formula provides much better accuracy compared to assuming the wavefunction vanishes outside of the boundaries (normal).}
  \label{fig:twofermions}
\end{center}
\end{figure}

For the $4$th order approximation, the second-derivative takes the form $\sim \frac{-30\psi_{0}+16\psi_{1}+16\psi_{-1}-\psi_{-2}-\psi_{2}}{12 \epsilon^2}$. Again, taking index $-1$ to refer to the left boundary, we see that the value of $\psi_{-2}$ is left unspecified. To maintain
the accuracy of the finite difference expression, we should choose $\psi_{-2}$ to smoothly continue the wavefunction past the boundary. In our case, we choose $\psi_{-2}=-\psi_{0}$ (i.e. the wavefunction is antisymmetric around the boundary). This means that at the boundary, the second-derivative should be replaced 
by $\frac{\partial^{2}\psi}{\partial x^{2}} \equiv \frac{29\psi_{0}+16\psi_{1}+\psi_{2}}{12 \epsilon^2}$. 
Continuing this argument, the coefficient $\frac{60}{12\epsilon^{2}}$ in Eq.~\ref{eq:discrete} is replaced
with different values depending on the nature of the boundary points:
at the red, blue and green points shown in Fig.~\ref{fig:Lattice}(b), the coefficients become $ \frac{58}{12\epsilon^{2}}$, $\frac{59}{12\epsilon^{2}}$ and $\frac{60}{12\epsilon^{2}} $. The final form of the $4$th order discretized lattice Hamiltonian thus becomes
\begin{eqnarray*}
 &&\mathcal{H}^{(4)}_{\epsilon}=\frac{16}{12\epsilon^{2}}\sum_{\langle ij\rangle,\sigma} c^{\dagger}_{i\sigma}c_{j\sigma}-\frac{1}{12\epsilon^{2}}\sum_{\langle \langle \langle ij\rangle \rangle \rangle,\sigma} c^{\dagger}_{i\sigma}c_{j\sigma}+h.c. \\
&&-\frac{60}{12\epsilon^{2}}\sum_{\sigma,i \in \text{red} }c^{\dagger}_{i\sigma}c_{i\sigma}-\frac{59}{12\epsilon^{2}}\sum_{\sigma,i \in \text{blue} }c^{\dagger}_{i\sigma}c_{i\sigma}-\frac{58}{12\epsilon^{2}}\sum_{\sigma,i \in \text{orange} }c^{\dagger}_{i\sigma}c_{i\sigma}
 \\&&+\bar{g}\sum_{i} c^{\dagger}_{i\uparrow}c_{i\uparrow}c^{\dagger}_{i\downarrow}c_{i\downarrow}-\mu \sum_{i\sigma}  c^{\dagger}_{i\sigma}c_{i\sigma}+\mathcal{O}(\epsilon^{4}),
\end{eqnarray*}
with $\epsilon=1/(N+1)$ and where the colors red, blue and orange correspond to the colored grid points in Fig.~\ref{fig:Lattice}(b).

To illustrate the importance of the correct representation of the Laplacian for
Dirichlet boundary conditions, in Fig.~\ref{fig:twofermions}
we show the relative error in the ground-state energy of two free fermions in the box, using different
treatments of the Laplacian, as a function of $\epsilon$. In Fig.~\ref{fig:twofermions}(a), we 
compare the lattice spacing $\epsilon = 1/(N+1)$ that is consistent with the boundary conditions to the
naive spacing $\epsilon=1/N$ for the Hamiltonian $\mathcal{H}^{(2)}$, showing that
the quadratic convergence in $\epsilon$ is achieved only for the former spacing. In Fig.~\ref{fig:twofermions}(b) we show the effect of using an antisymmetric continuation of the wavefunction in $\mathcal{H}^{(4)}$ compared to simply setting the value of the wave function outside of the box to zero; much faster convergence is obtained using the antisymmetric continuation.

\begin{figure}
\begin{center}
\includegraphics[width=1.0 \linewidth]{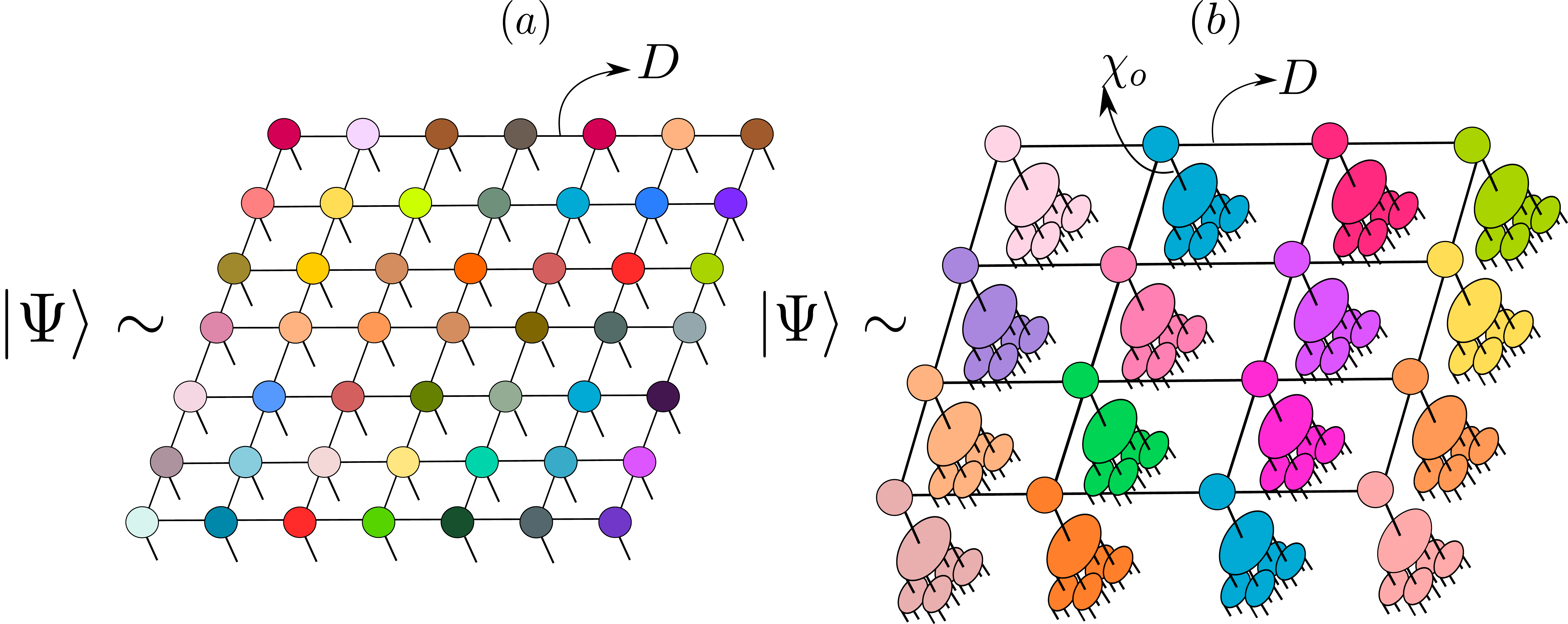} 
  \caption{(Color online) (a) The continuum limit wave function $|\Psi\rangle$ is approximated by a 
sequence of fPEPS of with bond dimension $D$ on successively finer lattices. The fPEPS on different lattices are related via a multigrid algorithm. This procedure is denoted fPEPS-fine. (b) The same continuum wave function $|\Psi\rangle$ can also be represented by a single fPEPS on a coarse lattice connected to layers of isometric tensors (here two layers are shown). We refer to this as fPEPS-tree.}
  \label{fig:peps}
\end{center}
\end{figure}

\section{Continuum fermionic tensor network ansatz}
\label{Sec:fermionic-tensor-network-ansatz}

We explore two different fermionic tensor networks to approach the continuum limit ground state
$|\Psi\rangle$, depicted in Fig.~\ref{fig:peps}(a, b). One (Fig.~\ref{fig:peps}(a)) is a standard fermionic PEPS (fPEPS) defined by a set of local tensors connected by virtual bonds corresponding to the geometry of the lattice discretization. The associated bond dimension of the virtual bonds is denoted by $D$ and controls the accuracy of the fPEPS ansatz. To enforce fermion statistics (i) all fPEPS local tensors are set to be symmetric under the action of $Z_2$ or $U(1)$ symmetry groups, and (ii) each line crossing in the network is replaced by a fermionic swap gate. Such an fPEPS captures fermionic states obeying an entanglement area law. As the lattice spacing goes to zero, the fPEPS then provides a numerical representation of the continuum limit. The primary numerical challenge is ensuring that the fPEPS tensors on the finest scales are properly optimized. This can be done by taking the numerical limit, i.e. connecting fPEPS representations at different discretizations, using a multi-grid algorithm discussed further below. We refer to this numerical continuum representation by fPEPS as fPEPS-fine. 


The second consists of several layers of isometric tensors, with a fPEPS placed at the top, see Fig.~\ref{fig:peps}(b). We denote this ansatz an fPEPS-tree. The isometric tensors are chosen to map four fermion sites onto one effective fermion site with bond dimension $\chi_{o}$. The isometries describe a coarse-graining transformation, where the parameter $\chi_{o}$ controls the accuracy
of the transformation. The amount of entanglement in the ansatz is controlled by the bond dimension of the topmost fPEPS, i.e. $D$, which encodes quantum entanglement between effective fermions on the coarsest level. The flexibility of the ansatz is thus controlled by both $\{\chi_{o}, D\}$. We expect this representation to work well in the dilute regime, where the effective area occupied by each fermion represents a coarse-grained length-scale and the isometries connect the finest (continuum) scale to that length-scale; in general, we expect $\chi_{o}, D \approx e^{\rho}$, where $\rho$ is the particle density. 
Although it is formally desirable, during coarse-graining, to decouple entanglement at each length-scale (as is the basis of fermionic multi-scale entanglement renormalization (fMERA))~\cite{Vidal:2008:MERA, Corboz:2009} the computational cost to work with fMERA is much higher than that of the fPEPS-tree. Thus, in fPEPS-tree
we account for the short-range entanglement entirely within the topmost fPEPS tensors.

\subsection{Tensor optimization and contraction techniques}

We now briefly summarize some of the techniques used to optimize and contract the different types of tensors appearing in the two ansatzes above. The fPEPS tensors are optimized towards a representation of the ground state using imaginary-time evolution, using the ``full-update'' method to perform bond truncations.~\cite{Corboz:2010:April, Lubasch:2014, Phien:2015} The full-update builds the environment from the entire wave function around each bond before truncation. We use the single-layer boundary contraction method\cite{Xie:2017} to contract
the fPEPS efficiently. The accuracy of the contraction is controlled by the boundary bond dimension $\chi_b$. Using these techniques, the computational cost of optimizing the fPEPS tensors is $\mathcal{O}(\chi_b^{3}D^{4})$. Assuming a boundary bond dimension $\chi_{b} \propto D^2$, this gives a computational cost of $\mathcal{O}(D^{10})$.

The isometric tensors in the fPEPS-tree are optimized using techniques similar to those used for the fMERA as described in Ref.~\onlinecite{Corboz:2009}. These are based on linearizing the respective cost functions with respect to the isometric tensors. In the fPEPS-tree, contracting a single layer of isometric tensors costs $\mathcal{O}(\chi_{o}^9)$. One advantage of the fPEPS-tree is that after a single isometric layer contraction, the fourth-order discretized Hamiltonian ($\mathcal{H}^{(4)}_{\epsilon}$) 
is renormalized into a nearest-neighbour Hamiltonian, which then retains its nearest neighbour form through subsequent isometric layers. This simplifies the optimization of the topmost fPEPS layer, which can be performed using standard nearest-neighbour imaginary-time evolution.

To improve efficiency, we exploit $Z_2$ and $U(1)$ symmetry in all tensors. We adopt the techniques developed in Refs.~\onlinecite{Singh:2011, Haghshenas:2018} to implement $U(1)$ symmetry, choosing relevant symmetric sectors during the optimization. We use a simple-update strategy (based on a direct SVD decomposition) to obtain an initial guess for the symmetry sectors, and those are then further dynamically updated during the full-update optimization by using a similar strategy.

\begin{figure}
\begin{center}
\includegraphics[width=1.0 \linewidth]{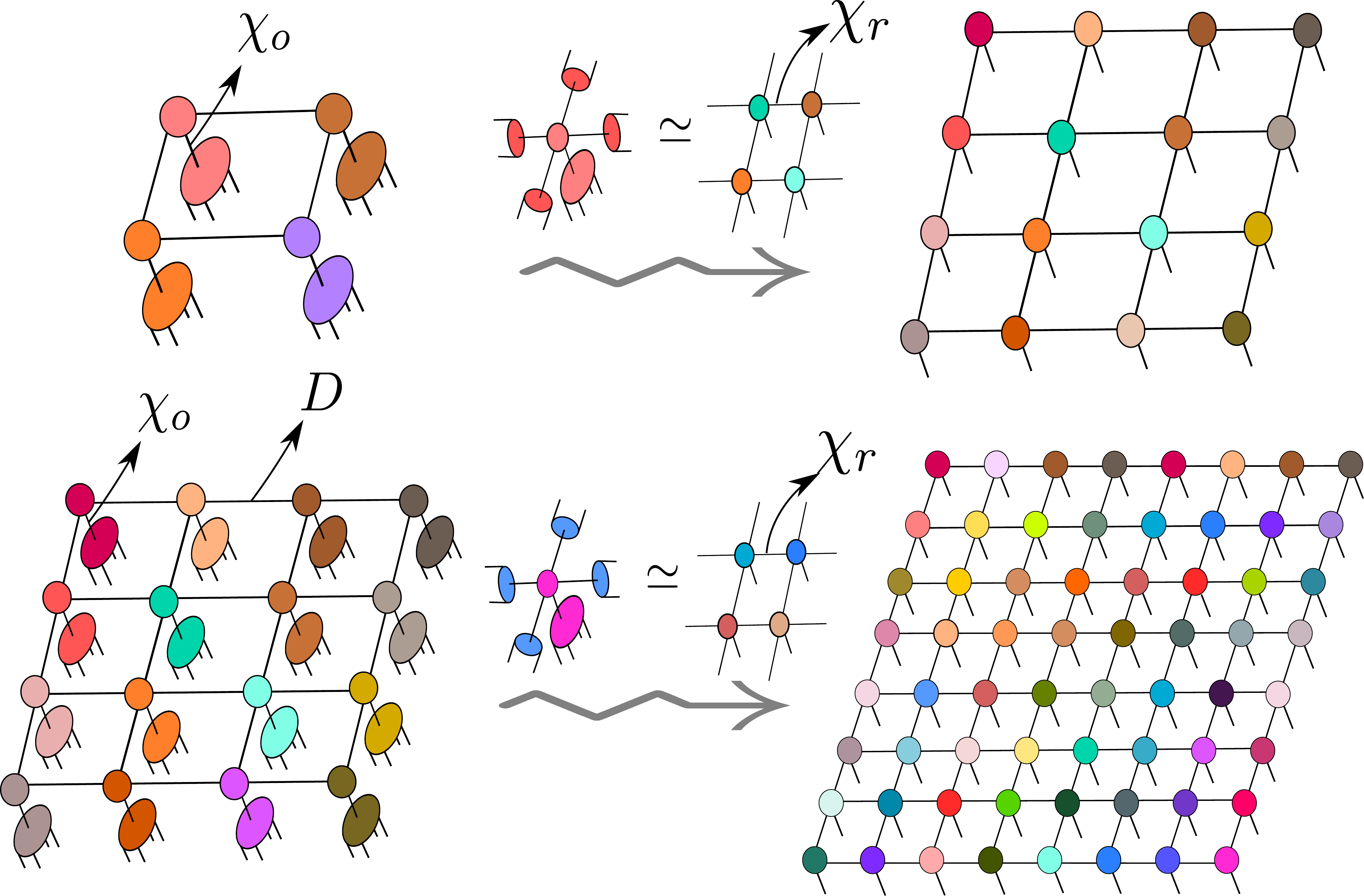} 
  \caption{(Color online) Diagrammatic representation of the multigrid algorithm. A splitting transformation is used to transform tensors from the coarser lattices to finer lattices. (a) A $4\times 4$ PEPS is obtained by using a splitting transformation for the coarser lattice. (b) By adding isometric tensors to the $4\times 4$ PEPS a good initial guess is provided for the tensors of the finer lattice, i.e. $8\times 8$. The procedure can be repeated to reach the desired fine lattice scale.}
  \label{fig:RG}
\end{center}
\end{figure}

\subsection{Multigrid fPEPS-fine optimization}

Although the fPEPS-fine wave function on the finest lattice is a straightforward representation of the (near)-continuum wave function,
direct optimization of such an fPEPS leads to numerical difficulties, such as slow convergence and being stuck in local minima.
This is analogous to what is seen in MPS simulations on very fine lattices~\cite{Dolfi:2012} and also what is seen
in solutions of partial differential equations on fine grids. Consequently, it is necessary to construct
the fPEPS on finer lattices from those on coarser lattices, which can be seen as taking the continuum limit
on the fPEPS tensors in an algorithmic sense. This we achieve using a multigrid-inspired algorithm.

\begin{figure}
\begin{center}
\includegraphics[width=1.0 \linewidth]{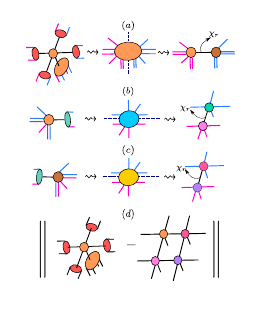} 
\caption{(Color online) Steps to construct the splitting map. (a) bonds with the same color are grouped together and an SVD is performed to split
  the tensor into two tensors. The truncation bond dimension is denoted $\chi_{r}$. (b, c) A resolution of the identity $U^{\dagger}U=\mathcal{I}$ is added to split a virtual bond to two virtual bonds. The green tensors denote $U$. After permutation, grouping bonds with the same color together, an SVD
  is performed. This procedure results in the final desired form. (d) We perform a direct optimization to further improve the accuracy of this transformation using the alternating least squares method.}
  \label{fig:RGSteps}
\end{center}
\end{figure}

The main idea in the multigrid approach is to interleave optimization and interpolation steps for the fPEPS tensors that are determined on lattices with different discretizations. In our version of the multigrid algorithm (i) we first approximate the ground state on the coarsest level by using a $N \times N$ fPEPS ansatz where $N \in \{2,3\}$, (ii) we attach a layer of isometric tensors to the $N \times N$ fPEPS to create an fPEPS-tree with a single layer of isometries for the $2N \times 2N$ lattice, and we subsequently perform energy optimization of the isometries and fPEPS tensors, (iii) we use a splitting map to map the fPEPS-tree to a $2N \times 2N$ fPEPS, then we relax the energy again on this finer lattice, (iv) we repeat steps (ii) and (iii) until the desired discretization level is reached, yielding the final fPEPS-fine wavefunction. A schematic of the multigrid algorithm is depicted in Fig.~\ref{fig:RG}. Note that the above is only one realization of a multigrid algorithm and many alternative choices can be made.

\begin{figure}
\begin{center}
\includegraphics[width=1.0 \linewidth]{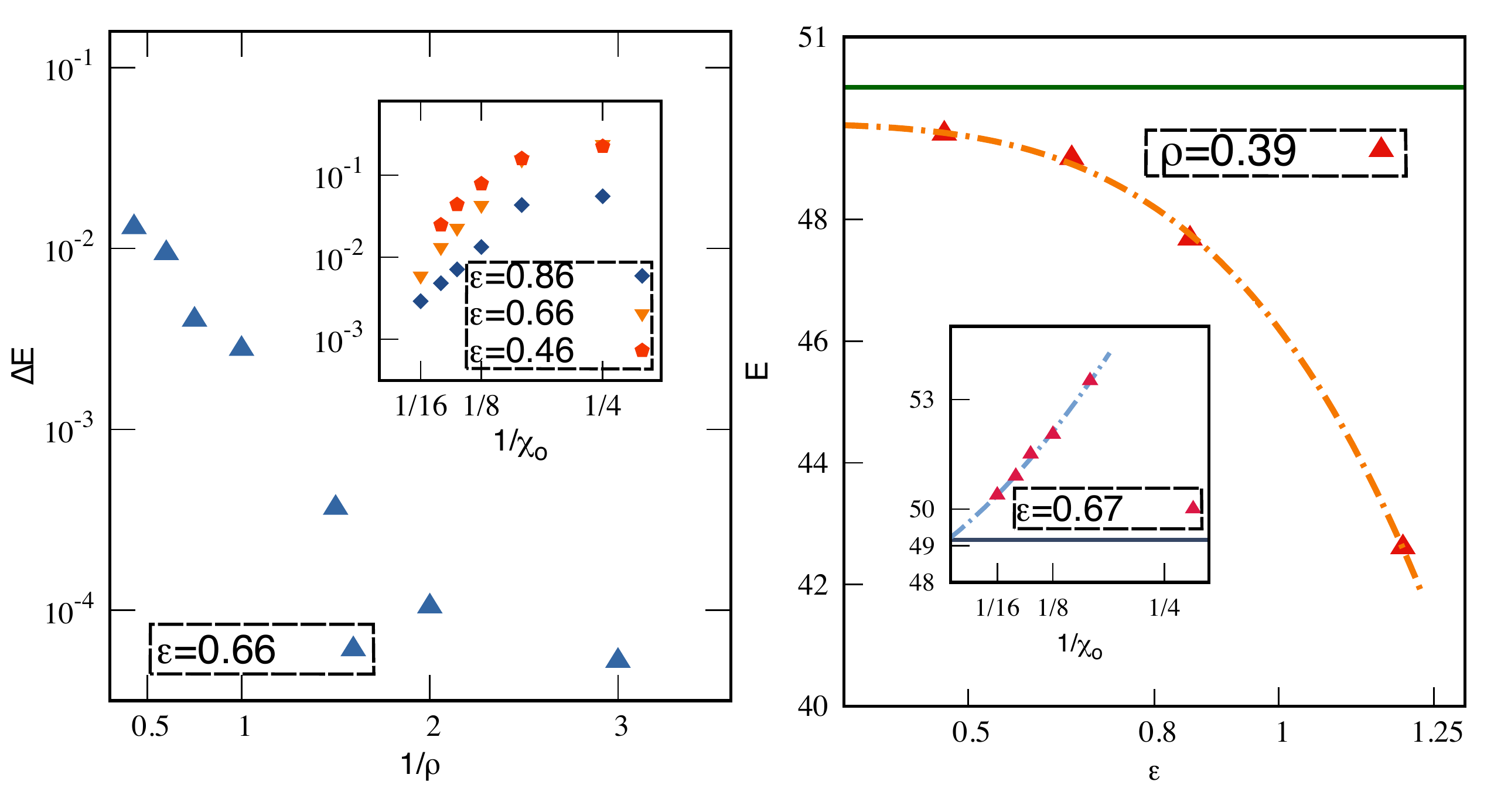} 
  \caption{(Color online) (a) The relative error of the free Fermi gas ground-state energy versus particle density using fPEPS-tree. The inset shows how the relative error behaves as a function of bond dimension $\chi_o$ and the lattice spacing $\epsilon$. (b) The ground-state energy versus lattice spacing $\epsilon$ for $\rho=0.39$. The inset shows the ground-state energy extrapolated as a function of bond dimension $1/\chi_o$ with fixed bond dimension $D=12$.}
  \label{fig:accrho}
\end{center}
\end{figure}

Two steps need to be further specified, namely (i) initializing the isometric tensors, and (ii) constructing
the splitting map. We initialize the isometric tensors by diagonalizing the local Hamiltonian, defined on the $2\times 2$ finer lattice, and picking the $\chi_{o}$ lowest-energy eigenvectors. This provides an approximate initial guess for the isometric tensors. The key steps in the splitting map are shown in Fig.~\ref{fig:RG}. We first add a resolution of the identity $U^{\dagger}U=\mathcal{I}$ onto the virtual bonds of the fPEPS, where $U$ is an isometric matrix with dimension $D^{2} \times D$. Such an isometric tensor $U$ (shown in red) splits one virtual bond into two virtual bonds, without changing the overall tensor network state. Then, one approximately solves the equation:  
\begin{equation} 
\label{EQ:SplittingTR}
\diagram{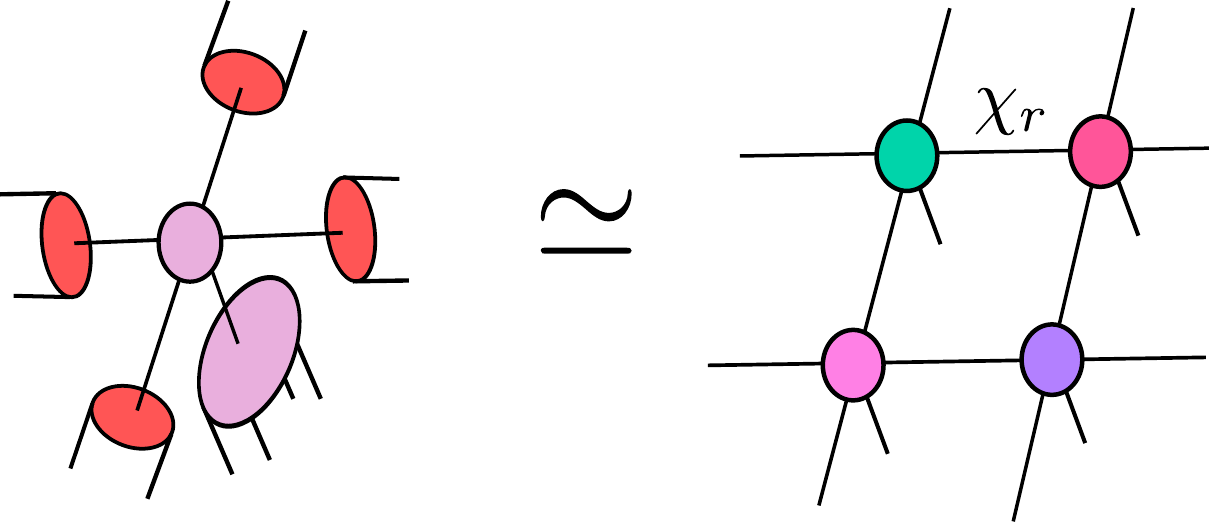}
\end{equation}
The parameter controlling the accuracy of this transformation is the bond dimension of the virtual bonds connecting the tensors, denoted  $\chi_{r}$; as $\chi_{r} \rightarrow \infty$ the transformation becomes exact. The tensors $U$ can in principle be considered to be variational parameters in order to best satisfy the above equation. However, in this paper, we fix the form of $U$; some of the diagonal elements are set to $1$, and the rest are set to $0$, i.e. $U_{ij, m}=\delta_{i\times D+j, m}$. This appears sufficient to obtain our desired accuracy (see Sec.~\ref{Sec:BenchmarkResults}). To solve Eq.~\ref{EQ:SplittingTR}, we carry out sequential SVD to obtain guesses for four resulting tensors, and then direct optimize the fidelity using alternating least squares to improve the accuracy of the splitting. In Fig.~\ref{fig:RGSteps}, we provide the details of this procedure. 

To summarize, the accuracy of calculations with fPEPS-fine using the multi-grid algorithm is controlled by four parameters:
the initial bond dimension $D$; the boundary bond dimension $\chi_{b}$ controlling the accuracy of the environment contractions; and the accuracy of the splitting map  controlled by parameters $\chi_{o}$ and $\chi_{r}$,
denoting the bond dimension of the isometries and the bond dimension of the resulting fPEPS on the finer lattice.
In practice, we can reasonably set parameters $\chi_b \sim D^2$ and $\chi_r \sim D $, thus the essential controlling parameters are only $D, \chi_o$.
As an example of the bond dimensions used in this paper, for the $16 \times16$ lattice fPEPS-fine simulation we used $(D,\chi_b)=(9, 200)$ and $(D,\chi_b)=(12, 250)$ for $Z_2$ and $U(1)$ symmetries, respectively.

\begin{figure}
\begin{center}
\includegraphics[width=1.0 \linewidth]{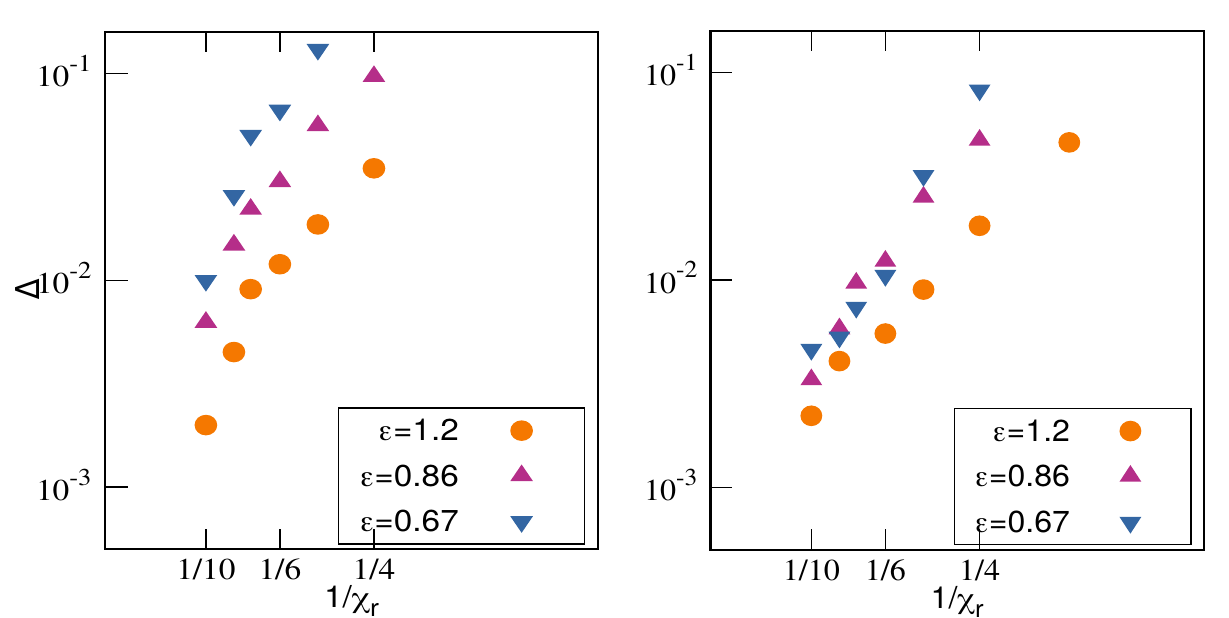} 
  \caption{(Color online) Relative error of the splitting map versus the bond dimension $\chi_{r}$ (a) for $\rho=0.17$ and (b) for $\rho=0.27$. The bond dimensions $(D, \chi_{o})$ are set to $(8,12)$. }
  \label{fig:splittingmap}
\end{center}
\end{figure}

\section{Results}
\label{Sec:BenchmarkResults}

\subsection{Free Fermi gas benchmarks}

We first assess the accuracy of the wave functions and algorithms discussed above using the free Fermi gas ($g=0$) as a benchmark system. This system is exactly solvable by reduction to single-particle quantities but is a challenging problem for tensor networks in the continuum and thermodynamic limit as the ground-state violates the entanglement area law by logarithmic terms, i.e. $\sim \rho^{\frac{1}{2}} \mathcal{A}\log{\mathcal{A}}$, where $\rho$ is the particle density and $\mathcal{A}$ is the boundary length.~\cite{Calabrese:2012, Vicari:2012} Consequently, to obtain accurate results, large bond dimensions in all steps of the algorithms are required
and approximation errors are magnified. In the calculations below, we shall use a fixed box side-length of $L=6$.


We first compute numerical continuum results using the fPEPS-tree ansatz using the $4$th order spinless Hamiltonian discretization. Here we use an fPEPS-tree with two layers of isometries. In Fig.~\ref{fig:accrho}(a), we show the relative error of the ground-state energy $\Delta E$ as a function of particle density $\rho$. As expected, the relative error increases sharply when we increase the particle density using fixed bond dimensions $(D, \chi_o)=(8,16)$. However, the inset shows that going to a finer lattice does not affect the accuracy significantly (the error increases only slightly in the continuum limit). In Fig.~\ref{fig:accrho}(b), we plot the ground-state energy $E$ versus the lattice spacing $\epsilon$, where we use the fourth-order polynomial function $E(\epsilon)=E_{\epsilon\rightarrow \infty }+b\epsilon^{-4}$ to extract the numerical continuum limit of the ground-state energy. Note that due to the use of coarse-graining, we no longer observe a simple $4$th order discretization error. For each data point for a specific lattice spacing $\epsilon$, we also perform a bond dimension extrapolation ($\chi_o \rightarrow \infty$). A second-order polynomial function $E(\chi_o)=E_{\chi_o\rightarrow \infty }+a \chi_o^{-1}+b\chi_o^{-2}$ is used to estimate the extrapolated results, as shown in the inset in Fig.~\ref{fig:accrho}(b). Because the energy is a function of the bond dimensions $(D, \chi_o, \chi_b)$,  to obtain a good extrapolated estimation, we need to make sure the results are converged with respect to the bond dimensions $(D,\chi_b)$. Thus at each $\chi_o$, we use as large $(D,\chi_b)$ as possible (up to $(12, 250)$). The extrapolated results are shown in Table~\ref{tab:fresult}. We find accuracies of roughly 1\% or better are obtained with these bond dimensions for the lowest 3 densities where $\rho < 0.4$.

\begin{figure}
\begin{center}
\includegraphics[width=1.0 \linewidth]{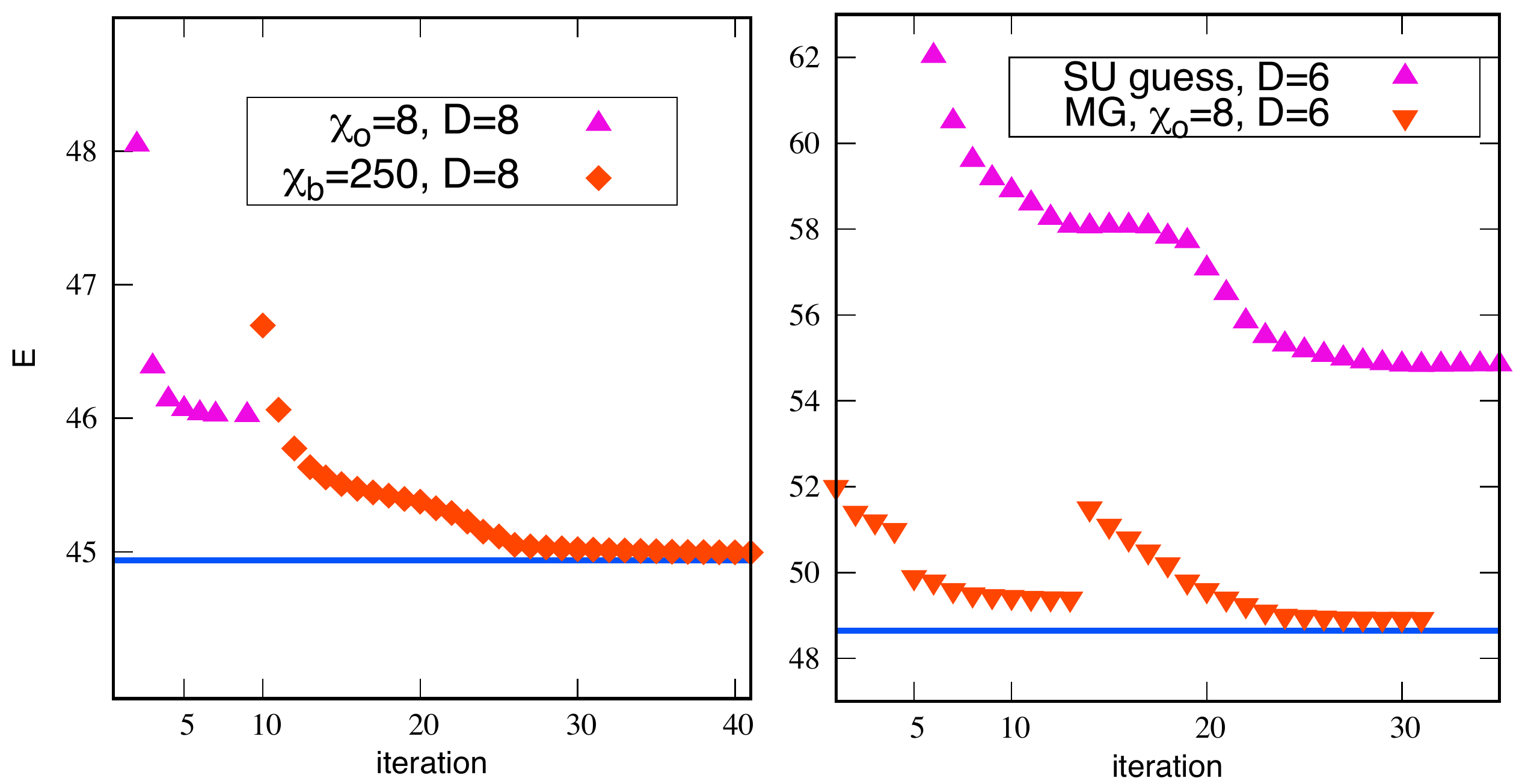} 
\caption{ (Color online) Behavior of the ground-state energy during the different steps of the multigrid algorithm. (a) The purple triangles denote the energies of the fPEPS ansatz after attaching isometries for $\rho=0.39$ and $\epsilon=0.67$. It provides an initial guess for the fPEPS ansatz on the finer lattice. The jump in the energies appears due to the approximate solution of the splitting equation Eq.~\ref{EQ:SplittingTR} with $\chi_r=8$. We observe that the fPEPS energies finally reach the exact result (blue solid line). (b) A comparison between the fPEPS energies optimized on the $16\times 16$ fine lattice, initialized by a simple update guess (labeled ``SU guess''), and by the multi-grid (MG) algorithm for $\rho=0.39$ and $\epsilon=0.35$. It is observed that the multi-grid algorithm provides a much better initial guess, hence better accuracy.}
  \label{fig:MGPEPS}
\end{center}
\end{figure}

We next explore the behaviour of the fPEPS-fine ansatz in the continuum free fermion problem.
As the multi-grid optimization involves several steps, we first illustrate the accuracy and errors associated with the individual steps. We start with the accuracy of the splitting map between coarser and finer lattices. This incurs an error which can be measured as $\Delta=\frac{E_b-E_f}{E_f}$, where $E_b$ and $E_f$ are the energies per lattice site before and after the splitting map, respectively. 

As $\chi_{r}$ increases we expect this error to go to zero, and for a practical method, it is important that a moderate $\chi_{r}\sim D$ is sufficient for good accuracy. In Fig.~\ref{fig:splittingmap}, we plot $\Delta$ versus the internal bond dimension $\chi_{r}$. We see that the relative error drops when increasing $\chi_r$ so that when $\chi_{r}\sim D$ it is  $\sim 10^{-2}-10^{-3}$. One might expect that if we proceed to finer lattice spacings $\epsilon$, (i.e. where the fine lattice has more sites), the relative error might increase due to the accumulation of individual tensor splitting errors. This is in fact seen in Fig.~\ref{fig:splittingmap}, although the relative error increases quite slowly with $\epsilon$. We also find that the accuracy of the splitting map does not depend on the particle density, as the relative error remains the same in the left and right panels. 

We further illustrate the numerical behavior of the multi-grid algorithm in Fig.~\ref{fig:MGPEPS}(a). As discussed above, applying isometries to the fPEPS to obtain a single-layer fPEPS-tree provides the initial guess for the splitting map (fPEPS-fine) with subsequent optimization being carried out after the splitting map is performed. We see that the ground-state energy jumps due to the infidelity of the splitting map, however, further optimization of the fPEPS on the finer level rapidly improves the energy. To show the importance of the multi-grid algorithm, in Fig.~\ref{fig:MGPEPS}(b), we compare the fPEPS energies initialized by a simple-update method~\cite{Jiang:2008} and the ones initialized from coarser lattices by the multi-grid algorithm. We observe that the fPEPS energies obtained by the multi-grid algorithm are much more accurate.

\begin{table}[t]
\begin{tabular}{|c|c|c|c|c|}
\hline $\rho$ &$\text{fPEPS-tree}, \mathcal{H}^{(4)}_{\epsilon}$ &$\text{fPEPS-fine}, \mathcal{H}^{(2)}_{\epsilon}$&$\text{exact}$  \\ \hline
$0.17$ &$10.95$&$10.99$&$10.966$  \\
$0.28$ &$27.34$&$27.5$ &$27.42$\\
$0.39$ &$49.6$ &$50.3$&$50.17$  \\ 
$0.55^\star$ &$50.8$&$54.0$& $54.8$  \\ \hline
\end{tabular}
\caption{Ground-state energy of the free Fermi gas for different particle densities in the continuum limit. Data with the symbol $\star$ are for spin-$\frac{1}{2}$ fermions, and data with no such symbol are for spinless fermions. Data
is better converged (and extrapolations are more accurate) at low density.}
\label{tab:fresult}
\end{table}

Using the above multigrid calculation with fPEPS-fine with up to four layers (a fine lattice of $16 \times 16$ sites) and the $2$nd order Hamiltonian discretization, we can estimate the energy of the fPEPS-fine ansatz in the numerical continuum limit. We use a linear extrapolation in $\frac{1}{D}$ (using a few largest values) to estimate the large-$D$ limit for each lattice spacing.\cite{Corboz:2016Jan} A polynomial function $E(\epsilon)=E_{\epsilon\rightarrow \infty }+b\epsilon^{-2}+c\epsilon^{-4}$ is used to estimate the numerical continuum limit for fPEPS-tree. The estimated energies are shown in Table~\ref{tab:fresult}. We see that in the very dilute regime, the fPEPS-tree is slightly more accurate than the multigrid algorithm, likely because it uses the $4$th order Hamiltonian discretization. However, in the non-dilute regime, the multigrid algorithm performs significantly better. Indeed for densities  $\rho<0.4$, the fPEPS-fine results are accurate to better than 0.4\%.

\begin{figure*}[ht!]
\centering
\includegraphics[width=1.0 \linewidth]{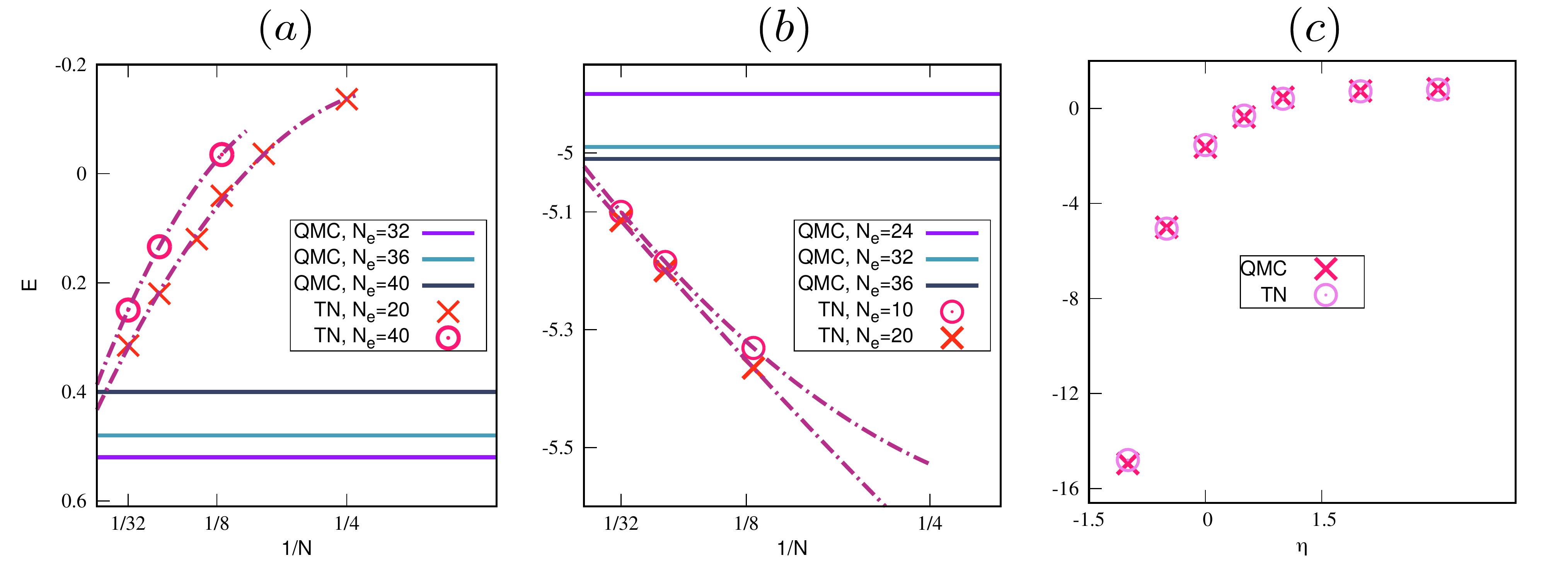}
\caption{ (Color online) Ground-state energy of the interacting 2D Fermi gas obtained  using the fPEPS-fine ansatz. $(a, b)$ The TN and QMC results for $\eta=\{1.0, -0.5\}$ versus lattice size $1/N$, respectively. The dashed lines show the inverse polynomial fitting function. $(c)$ Ground state-energy as a function of the dimensionless coupling parameter $\eta$. Note that the QMC
  data uses periodic boundary conditions while the TN data uses open boundary conditions, thus they only agree
in the thermodynamic limit.}
\label{fig:int}
\end{figure*}

\subsection{Interacting Fermi gas}
\label{sec:interactinggas}

We next present calculations using the fPEPS-fine ansatz for the spin-balanced interacting ($g < 0$) Fermi gas.
We focus on the spin-balanced regime because at this point auxiliary field quantum Monte Carlo has no sign problem,
and thus provides a reliable comparison; however, away from this point, sign problems manifest, for which the methods
developed here remain suitable.

Rather than using the interaction $g$ appearing in the continuum Hamiltonian, the
interaction strength is commonly parametrized using the dimensionless coupling parameter $\eta=\frac{1}{2} \log({2e_f}/{e_b})$,
where $e_f$ is the non-interacting Fermi energy and $e_b$ is the two-particle binding energy. We will be interested
in the simultaneous continuum and thermodynamic limit, which is obtained in the simultaneous limit of infinite particle number
and lattice size $N_e, N \to \infty$. For each finite lattice size $N$, an effective interaction $\bar{g}$ can be specified 
to be consistent with a given $\eta$, which defines the numerical fPEPS-fine lattice Hamiltonian.
The procedure to determine $\bar{g}$ is given in  Appendix B, and follows
that in Ref.~\onlinecite{Rammelm:2016}. 
The physics of the system is governed by $\eta$ and in the limits $\eta \ll 0$, $\eta \gg 0$, 
the system is in the Bose-Einstein condensate (BEC) and Bardeen-Cooper-Schrieffer (BCS) phases, respectively. Recent QMC studies have suggested that the BCS-BEC crossover occurs near $\eta\sim 1$.


We present fPEPS-fine results for the ground-state energy (for various $N$ and $N_e$) in units of the
non-interacting Fermi energy $E=\frac{\langle \mathcal{H} \rangle}{N_e e_f/2}$. 
$\mathcal{E}=\lim_{N_e, N \rightarrow \infty} E$ is the desired continuum and thermodynamic limit result. We compare to results from auxiliary field QMC (AFQMC) which is exact up to statistical error (for given $N_e, N$). The AFQMC data is computed using periodic boundary conditions whereas the fPEPS-fine energies are computed for open boundary conditions, however, in the thermodynamic limit they should approach the same value. In Fig.~\ref{fig:int}(a, b), the ground-state TN results as a function of $N_e, N$ are compared against the AFQMC data (shown as lines, extrapolated to the thermodynamic limit for each $N_e$). To extrapolate the TN data to the thermodynamic limit, 
we use the second-order polynomial function $E = E_{N \rightarrow \infty}+bN^{-1}+cN^{-2}$. For the point $\eta=-0.5$ (Fig.~\ref{fig:int}(b)), the thermodynamic and continuum limits are rapidly approached as seen in both the TN and AFQMC data; however, this is more challenging for the crossover point $\eta=1.0$ (Fig.~\ref{fig:int}(b)) where there are sizable finite-size effects in both the TN and AFQMC results. However, by using up to $N_e=40$ particles, and lattices with up to $32\times 32$ sites, the continuum TN data and AFQMC extrapolations provide consistent estimates.


We further show the extrapolated continuum and thermodynamic limit fPEPS-fine and AFQMC ground-state energies across a range of renormalized coupling parameters, in Fig.~\ref{fig:int}(c). We find good agreement between the fPEPS-fine and AFQMC energies, with the largest errors around
the transition point $\eta \sim 1.0 $, which again comes mainly from the uncertainty in the $N_e, N\to \infty$ extrapolations required in both methods.

\section{Conclusions}
\label{Sec:CONCLUSION}
We have described the numerical continuum limit of two-dimensional
tensor network states based on two variants of projected entangled pair states, as well as the numerical algorithms
used to work with them. Using continuum grids with up to approximately 1000 sites,
our initial calculations show promising results  for two fermionic continuum systems in two dimensions: the entanglement law violating free Fermi gas,
as well as the interacting unitary Fermi gas of much interest in cold atom experiments. In the latter case, our results
compare well to auxiliary field quantum Monte Carlo calculations that are feasible at the spin-balanced point.
However, the strength of the continuum tensor network approach is that it is not limited to special points
in the phase diagram. This opens up the use of tensor networks to address  unresolved questions in spin-polarized Fermi gases,~\cite{Bulgac:2006, Bulgac:2007, Randeria:2014}
as well as in other problem areas, such as the realistic description of electronic structure with tensor networks, and the numerical study of
field theories in two-dimensions and higher.

\acknowledgments
This work was supported by the US National Science Foundation (NSF) via grant CHE-1665333. GKC acknowledges support from the Simons Foundation via
the Many-Electron Collaboration and via the Simons Investigator program. We have used the  \emph{Uni10} library\cite{Kao:2015} for
implementation and \emph{PySCF} \cite{Qiming:2017} to obtain benchmark data. We thank H. Shi, J. Drut, and T. Berkelbach for helpful discussions
regarding the unitary Fermi gas.

\bibliography{Ref}


\appendix

%
 
\section{Discretization of the unitary Fermi gas}
\label{sec:appendixb}
In the continuum limit, given an attractive interaction ($g <0$) two particles will have a bound ground-state under the continuum
Hamiltonian in Eq.~\ref{eq:Ham}. We can therefore use the binding energy $\epsilon_b$ as a measure of the strength of this interaction.
In the thermodynamic limit, the other parameter required to specify the state is the density $\rho$. This is reflected
in the Fermi energy $\epsilon_f = k_f^2/2$ via $k_f = \sqrt{2\pi \rho}$. Thus we can characterize the physics of the system
via the dimensionless ratio $\epsilon_f/\epsilon_b$, or equivalently $\eta=\frac{1}{2} \log({2\epsilon_f}/{\epsilon_b})$.

In a discretized version of the problem, we can imagine a box of sidelength $N \epsilon$, then $\rho = N_e/N^2 \epsilon^2 = n/\epsilon^2$
and $\epsilon_f = \pi N_e/N^2 \epsilon^2 = e_f/\epsilon^2$. We can also write $e_b = \epsilon_b/\epsilon^2$, where
$e_b$ is the binding energy of the lattice Hamiltonian (e.g. in the $2$nd order discretization)
\begin{eqnarray*}
{H}^{(2)}&&=\sum_{\langle ij\rangle} c^{\dagger}_{i\sigma}c_{j\sigma}+h.c.-4\sum_{i}c^{\dagger}_{i\sigma}c_{i\sigma}\\
&&+\epsilon {g}\sum_{i} \mu_i c^{\dagger}_{i\uparrow}c_{i\uparrow}c^{\dagger}_{i\downarrow}c_{i\downarrow}-\mu \sum_{i}  c^{\dagger}_{i\sigma}c_{i\sigma}
\end{eqnarray*}
Note that a given ratio $\epsilon_f/\epsilon_b$ fixes the same ratio $e_f/e_b$, thus the lattice spacing $\epsilon$ drops out except via the effective coupling $\bar{g}=\epsilon{g}$. However, since the functional relationship between $g$ and $\epsilon_b$ is not known a priori, we can simply adjust $\bar{g}$ to obtain the desired $e_b$ by solving for the two-particle binding energy on a lattice of side-length $N$. Thus the discretization parameter does not appear explicitly.

\end{document}